# Impact of internal migration on population redistribution in Europe:
# Urbanisation, counterurbanisation or spatial equilibrium?

**Francisco Rowe** [1*], **Martin Bell** [2,3], **Aude Bernard** [2,3], **Elin Charles-Edwards** [2,3] **and Philipp Ueffing** [4]

[1] *Geographic Data Science Lab, University of Liverpool, Liverpool, United Kingdom*
[2] *University of Queensland, Brisbane, Australia*
[3] *Shanghai University, Shanghai, China*
[4] *United Nations Population Division, New York, United States*

*Corresponding author: Francisco Rowe (F.Rowe-Gonzalez@liverpool.ac.uk)

**Abstract**
The classical foundations of migration research date from the 1880s with Ravenstein's 'Laws of migration', which represent the first comparative analyses of internal migration. While his observations remain largely valid, the ensuing century has seen considerable progress in data collection practices and methods of analysis, which in turn has permitted theoretical advances in understanding the role of migration in population redistribution. Coupling the extensive range of migration data now available with these recent theoretical and methodological advances, we endeavour to advance beyond Ravenstein's understanding by examining the direction of population redistribution and comparing the impact of internal migration on patterns of human settlement in 27 European countries. Results show that the overall redistributive impact of internal migration is low in most European countries but the mechanisms differ across the continent. In Southern and Eastern Europe migration effectiveness is above average but is offset by low migration intensities, whereas in Northern and Western Europe high intensities are absorbed in reciprocal flows resulting in low migration effectiveness. About half the European countries are experiencing a process of concentration toward urbanised regions, particularly in Northern, Central and Eastern Europe, whereas countries in the West and South are undergoing a process of population deconcentration. These results suggest that population deconcentration is now more common than it was in the 1990s when counterurbanisation was limited to Western Europe. The results show that 130 years on, Ravenstein's law of migration streams and counter-streams remains a central facet of migration dynamics, while underlining the importance of simple yet robust indices for the spatial analysis of migration.

1. **Introduction**

The classical foundations of migration research date from the late 19th Century with Ravenstein's 'Laws of migration' (Ravenstein 1885), which represent the first systematic comparative analyses of internal migration. Drawing on empirical regularities from the 1871 and 1881 British census, Ravenstein proposed seven laws, which have been empirically validated, except perhaps for the observation that urban dwellers are less mobile than their rural counterparts, which reflects the time when Ravenstein wrote. While Ravenstein's propositions have been criticised as descriptive, deterministic and historical (Castles/Miller 1993), they '*provided the hypotheses upon which much future migration research and theorisation was built*' (Boyle et al. 1998: 5), including gravity and human capital models (Greenwood 2019). Ravenstein's theoretical propositions are simple, yet broad ranging as they are concerned with several different aspects of migration behaviour and address a range of questions that remain of relevance today: who migrates? why do they migrate? where do they migrate? and how often do they migrate?

Ravenstein is also credited with the first cross-national comparison of internal migration. His 1889 paper (Ravenstein 1889) extended the search for empirical regularities to over 20 European countries, Canada and the United States by examining lifetime net migration aggregates, concluding that migratory movements follow the same principles in all countries. Ravenstein wrote at a time of rapid industrialisation when rural-to-urban flows dominated migration systems in Europe. Yet, in reference to his fourth law, Ravenstein observed that for '*each main stream or current of migrants there runs a counter-current, which more or less compensates for the losses sustained by emigration. This counter-current is strong in some cases, weak in others, and literally compensatory in a few instances*' (1885: 187). Like many subsequent observers (Zelinsky 1971), Ravenstein did not anticipate the changes in direction of the net balance between urban and rural flows, which led to counterurbanisation in Western Europe in the 1970s (Champion 1989; Fielding 1989). Nevertheless, his observations encompass all the key elements for understanding the dynamics of contemporary migration flows.

The 20th century has seen considerable progress in data collection practice and methods of analysis, and a rich comparative literature has progressively developed in Europe focussing on the intensity, composition, patterning and spatial impacts of population movement (Champion 1989; Fielding 1989; Rees/Kupiszewski 1999; Rogers/Castro 1983; Rowe 2018a). Despite these contributions, contemporary understanding of the way migration impacts on settlement patterns remains crude, constrained primarily by reliance on coarse dichotomies into urban and rural, data inadequacies and the perennial obstacles presented by the modifiable areal unit problem (MAUP) (Openshaw 1975). These difficulties are compounded when seeking to make comparisons between countries and to trace the changing effect of migration on human settlement patterns over time.

Recent analytical and theoretical advances developed as part of the Internal Migration Around the GlobE (IMAGE) project now provide the means to circumvent these difficulties. Moving beyond the urban-rural dichotomy, Rees et al. (2017) advanced a theoretical framework which captures the way internal migration redistributes population across the national settlement system during the development process. They also proposed system-wide measures of internal migration impact which are independent of the size and number of spatial units used for measurement and identified systematic links between migration intensity and migration effectiveness. Coupled with a global repository of internal migration data (Bell et al. 2015a) and bespoke software, the IMAGE Studio (Stillwell et al. 2014), these developments provide the framework and tools to systematically quantify the impact of internal migration on population redistribution within countries, and explore cross-national differences.

Drawing on the above resources, we compare the impact of internal migration on patterns of human settlement across 27 European countries; determine the direction and pace of population concentration due to internal migration within each country; and examine changes in migration impact over time. In doing so, we aim to assess the relevance of Ravenstein's fourth law of streams and counter-streams to contemporary migration systems. By focusing on Europe, we seek to update and extend the work of Rees/Kupiszewski (Rees/Kupiszewski 1999) who explored cross-national variations in internal migration patterns across 11 European countries between the mid-1980s and mid-1990s. Profound demographic and economic changes have occurred at the start of the new millennium. Open borders and Eurozone expansion, coupled with a European debt crisis and the acceleration of population ageing, are likely to have altered the patterns of internal migration within European countries by facilitating international migration and increasing socio-economic inequalities between sub-national areas (De Beer et al. 2010; Dubois et al. 2007; Ertur et al. 2006). International migration may have operated to substitute for within-country moves (De Beer et al. 2010), while growing sub-national disparities may have redirected the concentration of internal population flows towards a handful of destinations (Dubois et al. 2007; Ertur et al. 2006). The recent and changing nature of these events underpins a need for an update, review and monitoring of internal migration trends.

The paper is structured in five sections. In Section 2, we review prior work and highlight the extent of cross-national variations in the spatial impact of internal migration. In Section 3, we discuss impediments to cross-national comparison arising from differences in data types, observation intervals and geographical frameworks. In subsequent sections, we present the results of our analysis which proceeds in a series of stages. In Section 4, we first assess the overall impact of internal migration on population redistribution using the Index of Net Migration Impact (INMI), a single system-wide index which transcends national differences in the zonal systems on which migration is recorded. We compare the level of redistribution in European countries to the world average and examine how cross-national differences are driven by the interaction between migration intensity and migration effectiveness. In Section 5, we examine how these system-wide differences play out to alter the pattern of human settlement at the local and regional level, moving beyond conventional measures based on the urban hierarchy to identify the overall effects on population concentration and deconcentration within countries, highlighting unusual patterns of population gains and losses. To that end, we set net migration rates against population densities across entire national zonal systems and compare the slope of population-weighted regressions for our sample of countries. In Section 6, we examine trends over time in the context of the conceptual model proposed by Rees et al. (2017) which anticipates a range of trajectories among economically advanced countries. Section 7 provides concluding remarks by discussing the long-standing differences in migration processes that distinguish the different regions of Europe and identify the need for a more comprehensive view of population movement that recognises the multi-dimensional nature of the migration process.

## 2. Prior Work

Compared with other parts of the world, the level, characteristics and impact of internal migration in many countries of Europe are well understood. This is due to sustained research by demographers and geographers but also reflects good data availability, with records on internal migration in parts of northern European stretching back centuries. Comparative studies are less common, reflecting the difficulties in harmonising migration data derived from different sources, over varying time intervals, and for different statistical geographies (Bell et al. 2015a). A number of cross-national studies have sought to overcome these challenges, to enumerate variations in intensity and the age profile of migration, and to assess its spatial impact across the continent.

Cross-national variation in the intensity, or level of migration, is arguably the best-understood dimension of migration but is challenging to establish due to its sensitivity both to the interval over which migration is measured and to the size of the areal units used to define migration. Early cross-national studies of internal migration (Parish 1973; Rogers et al. 1983) were severely hampered by these issues. Parish (1973) circumvented the problem by comparing trends, rather than the absolute level of migration, for eight countries in the 19th and early 20th Century. The study found modest increases in intensity in all countries, though both the timing and magnitude of the increase varied, highlighting the importance of local context. The International Institute for Applied Systems Analysis' (IIASA) comparative study of migration and settlement was the first to capture regional variations in internal migration for a large sample of countries, including 13 countries in Europe. An index capturing the level of retention within regions (Rogers et al. 1983) as well as gross migraproduction rates (Rogers/Castro 1983) were calculated. Median values for regions were used as a comparative measure of differences in regional migration intensity between countries. Migration intensity was lowest in Bulgaria and Austria, while high levels of mobility were recorded in regions of Hungary, Finland and the former members of the Soviet Union. Differences in areal delineation affected the reliability of results, so the authors focused instead on the age profile of migration (Rogers/Castro 1983).

Rees/Kupiszewski (1999) were the first to apply a robust single measure of migration intensity, Courgeau's k (Courgeau 1973), for the purposes of cross-national comparison. Their analysis of 11 countries in Europe revealed a gradient of high mobility in northern and western Europe tending to low mobility in southern and eastern Europe. Sanchez/Andrews (2011) using consistent data from the 2007 European Union Survey of Income and Living Conditions, confirmed this general pattern, as did Bell et al. (2015b) who estimated a novel, system-wide index, the aggregate crude migration intensity (ACMI) for 96 countries around the world including 30 countries in Europe. Esipova et al. (Esipova et al. 2013) confirmed the northwest-southeast gradient using data from a standard question in the 2011-2012 Gallup World survey. More recently, Bernard (2017) adopted a cohort perspective to compare the lifetime number of moves of early baby boomers in 14 European countries, and Rowe (2018a) examined short- and long-distance migration patterns for a sample of 27 European countries. Both confirmed a clear spatial gradient of high mobility in the North and West, moderating toward the South and the East. Champion et al. (2018) provide a comprehensive account of trends in internal migration intensities in the developed world. This work has revealed for the first time a diversity of trends in internal migration in Europe, with some countries recording a decrease while others show stability or increase (Bell et al. 2018).

The most visible significance of internal migration lies in its effect in redistributing populations, which is heightened under conditions of low fertility and mortality. Comparative studies of migration impact in Europe stretch back decades, many focusing on the contribution of migration to urbanisation. Fielding (1989) explored changes in the direction of population redistribution processes in 14 European countries. Population density was used as a proxy for the level of urbanisation in regions and correlated against regional net migration rates. The results revealed a shift in the dominant spatial impact of migration away from urbanisation to counterurbanisation between the 1950s to the 1970s.

However, these shifts were neither linear nor ubiquitous and the lack of a comparable summary metric prohibited rigorous comparison of the relative magnitude of population redistribution. Champion/Vandermotten (1997) drew on estimates of net migration for 557 regions of Europe in the three decades from the 1960s. Migration was measured as a residual after subtracting natural change (births-deaths) and therefore did not isolate the impact of internal migration on regional populations. This study found strong associations between population density and regional migration gains in the 1970s. However, the relationship was non-linear, with the largest gains recorded in regions with intermediate population densities. In the 1960s and 1980s, regional economic factors accounted for more variance in net migration rates than density, pointing to equilibrating economic forces sitting alongside structural drivers of migration in Europe (Champion/Vandermotten 1997).

Rees/Kupiszewski (1999) building on this earlier work examined the association between urbanisation (again adopting population density as a proxy) and patterns of regional net migration gains and losses. The study identified three main systems of population redistribution in Europe: urbanisation (Estonia, Romania, Norway, and Poland); intermediate systems (Germany, Italy,

Portugal and the Czech Republic) and counter-urbanising systems (Netherlands and the United Kingdom). Net migration gains and losses were also impacted by regional unemployment in certain countries. Unemployment was strongly associated with net losses in the United Kingdom and Poland, while Portugal, Italy, and the Czech Republic displayed weak associations. The selectivity of migration with respect to gender and life course stage was also examined, with evidence of differences in the pattern of population redistribution according to the characteristics of migrants especially with respect to life course stage.

The cumulative knowledge gained over previous decades, set alongside recent findings from the IMAGE project (Bell et al. 2015b; Stillwell et al. 2016), provides distinctive insights into cross-national variations in internal migration in Europe. Compared with other parts of the world, the overall intensity of migration is moderate, but variable across countries (Rowe 2018b). Multiple studies have found evidence of a spatial gradient of high mobility in Northern and Western Europe to low mobility in Southern and Eastern Europe (Bell et al. 2015b; Rees/Kupiszewski 1999; Rowe 2018a; Sánchez/Andrews 2011). The impact of migration on settlement systems also varies across the continent and over time, but with less clarity in spatial patterning. Both urbanising and counter-urbanising tendencies are evident. This is overlain by systems of migration flows reflecting the relative economic fortune and function of regions. A lack of comparable system-wide metrics has made the overall impact of migration difficult to quantify and compare across countries.

### 3. Internal Migration Data in Europe

In a recent global review, Bell et al. (2015a) found that 41 of 43 European countries collected data on internal migration. Europe is unusual in making significant use of population registers but a roughly equal number of countries drew on registers, censuses and surveys as their principal source of internal migration data, with fully 34 countries using more than one source. Differences in collection instruments hinder comparability between countries because registers and censuses measure migration in different ways. Population registers record migration events, whereas censuses measure transitions between discrete points in time, therefore counting migrants, rather than migrations (Rees et al. 2000). When measured over lengthy intervals, event and transition data provide a different picture of internal migration because registers record multiple moves, which transition data fail to capture. Over short intervals, however, such as one year, event and transition data tend to deliver similar results, with the number of migrants closely matching the number of migrations (Long/Boertlein 1990). Fortunately, census collections in most European countries measure migration over a single year interval, which corresponds with the data commonly available from population registers.

Drawing on migration data from the IMAGE repository (Bell et al. 2015a), we use information from both population registers and censuses to maximise geographical coverage. Our dataset comprises the 27 countries listed in Table 1, which accounts for 98.6% of Europe's population. Excluded are twelve small city states and countries with small populations or for which no data were readily available: Albania, Andorra, Bosnia-Herzegovina, Iceland, Lichtenstein, Luxembourg, Macedonia, Moldova, Monaco, Montenegro, San Marino and Serbia. Also excluded are five countries in which migration is measured over less than 20 administrative units: Cyprus, Latvia, Malta, Slovakia and Slovenia. Since Turkey, like Russia, straddles the continental divide, we also include them in our dataset. For 20 of our 27 countries, data were drawn from population registers or administrative sources, with the remaining seven coming from censuses. Of the latter, just two, France and Switzerland, collect data over a five-year transition interval, which calls for some care when making cross-national comparisons. In the analyses which follow, we address this by comparing the results for Europe against the global mean reported by Rees et al. (2017) for which a larger sample of countries is available.

Comparability between the 27 countries in our dataset is compromised by two further issues: differences in the years for which migration was observed and variations in the spatial scale at which it was measured. For the former, we chose to balance the goals of comparability and need for up-to-date data, and elected to use data for the latest period available in the IMAGE repository. Although the timing of observation varies across a full decade from 2000-02 (e.g. Switzerland and Romania) to 2012-13 (e.g. Norway), the data in Table 1 provide a broad representation of migration patterns at the start of the millennium. As demonstrated below, they also reveal wide variations between countries in the direction and extent of redistribution, although we sometimes find substantial variations in these trajectories from year to year. We hold origin-destination matrices for most countries, with the exception of Hungary for which we only have in-migration and out-migration flows for each administrative unit separately. Similarly for Bulgaria we hold in-migration and out-migration flows for 264 municipalities but an origin-destination matrix only at a provincial level (n=28). While inflows and outflows are sufficient to estimate net migration rates in Section 5, a full matrix is required to estimate measures used in Section 4, which causes the number of case study countries to vary across the paper.

**Table 1**. List of country, data types, years and number of regions.

| Country | Year | Data type | No. of Regions |
|---|---|---|---|
| Austria | 2010 | Event | 99 |
| Belarus | 2011 | Event | 130 |
| Belgium | 2005 | Event | 589 |
| Bulgaria* | 2006 | Event | 28/264 |
| Czech Republic | 2010 | Event | 77 |
| Denmark | 2011 | Event | 99 |
| Estonia | 2010 | Event | 225 |
| Finland | 2011 | Event | 336 |
| France** | 2006 | Transition | 22 |
| Germany | 2009 | Event | 412 |
| Greece | 2011 | Transition | 54 |
| Hungary*** | 2010 | Event | 196 |
| Ireland | 2006 | Transition | 26 |
| Italy | 2009 | Event | 107 |
| Lithuania | 2010 | Event | 60 |
| Netherlands | 2010 | Event | 431 |
| Norway | 2013 | Event | 428 |
| Poland | 2010 | Event | 379 |
| Portugal | 2011 | Transition | 30 |
| Romania | 2002 | Event | 42 |
| Russia | 2010 | Event | 80 |
| Spain | 2011 | Transition | 52 |
| Sweden | 2012 | Event | 290 |
| Switzerland** | 2000 | Transition | 184 |
| Turkey | 2012 | Transition | 81 |
| Ukraine | 2010 | Event | 27 |
| United Kingdom | 2011 | Transition | 404 |

Note: * Origin-destination matrix between 28 provinces and in-migration and out-migration flow for 264 municipalities; **Five-year transition data; *** in-migration and out-migration flows.

Comparison of migration patterns based on varying time points may be complicated by the fact that countries follow different trends. The period of analysis encompasses the 2007/08 global financial crisis (GFC) and recent studies have attributed changes in the level of migration to this discrete economic event (Bell et al. 2018; Lomax/Stillwell 2018). Yet, changes in migration intensity vary across countries. Analysing a time series of annual change in migration rates for a period of up to 39 years across 27 European countries, Rowe (2018a, 2018b) revealed that only five countries -Iceland, Bulgaria, Romania, Latvia, Belarus and Croatia- experienced a pronounced decline during the GFC period between 2007 and 2009. The predominant pattern across countries across most countries was of cyclical fluctuations in the long-term trajectory of migration intensity. Yet, these studies have focused on the level of migration, rather than on the direction of migration flows. The impacts of the GFC on influencing the direction of migration flows is difficult to anticipate and expected to vary across countries according to the national structure of population distribution and local economic conditions.

The second issue that affects comparability arises from variations in the zonal systems across which migration is measured. These zonal systems range in our dataset from less than 30 regions in Bulgaria and France to more than 500 in Belgium. Differences in the number of regions into which a country is divided, and in their shape and size, fundamentally affect the number of migrations captured, and therefore influence any derived statistics. This is widely recognised as the MAUP which affects spatial modelling (Openshaw 1975). Courgeau et al (2013) showed that this problem could be overcome by setting migration intensities against the average number of households per zone at a range of spatial scales, to derive an estimate of the aggregate crude migration intensity (ACMI), a measure of all moves within each country, irrespective of distance moved, which was directly comparable across nations. Compared with the measure devised four decades earlier by Courgeau (1973), latterly referred to as Courgeau's k (Bell/Muhidin 2011, 2009; Rees/Kupiszewski 1999), it offered the distinct advantage of having an intrinsic meaning. Stillwell et al. (2014) describe the random spatial aggregation software implemented as part of the IMAGE project to provide a general solution for countries with migration data available on a finely grained spatial framework. Rees et al. (2017) subsequently used these routines in the IMAGE Studio to measure the scale and pattern effects of the MAUP on two key indicators of migration impact: the Aggregate Net Migration Rate (ANMR) and the Migration Effectiveness Index (MEI). Coupling these measures with the ACMI, they derived a new index, the Index of Net Migration Impact (INMI), as used in this paper, which allows robust, system-wide comparisons between countries in regard to migration redistribution.

## 4. Overall Impact of Internal Migration

The INMI is a generalised extension of the ANMR originally proposed as one of the key measures of migration impact by Bell et al. (2002). Algebraically, the ANMR is defined as half the sum of the absolute net changes across all regions, divided by the population at risk $P$:

$$ANMR = 100 * 0.5 \sum_i |D_i - O_i|/P \qquad (1)$$

where $D_i$ and $O_i$ represent in-migration and out-migration flows from region $i$. Bell et al. (2002) also show that the ANMR is the product of two other key migration measures, the crude migration intensity (CMI) and the migration effectiveness index (MEI):

$$ANMR = CMI * MEI/100 \qquad (2)$$

where:

$$CMI = 100M/P \qquad (3)$$
$$MEI = 100 * 0.5 \sum_i |D_i - O_i|/M \qquad (4)$$

And M indicates the total number of inter-regional migrants. The CMI measures the overall level or incidence of migration within a country, whereas the MEI indicates the degree of symmetry or balance between migration inflows and outflows – a system-wide measure of Ravenstein's observation with regard to the reciprocal nature of migration flows. While providing a robust and informative index of population redistribution in a single country, values of the ANMR are clearly dependent on spatial scale. Since ANMRs calculated for differing levels of geography are clearly not comparable, the index is unsuitable for cross-national comparisons. Rees et al. (2017) devised a general solution by using the random spatial aggregation facility in the IMAGE Studio (Stillwell et al. 2014) to assess the effects of the MAUP on the CMI and the MEI. While the CMI was found to increase linearly with the log of the number of spatial units (Courgeau 1973), the MEI tends to remain remarkably stable when calculated for geographies of 20 spatial units or more. Note that the CMI differs from the ACMI which, as explained earlier, is scale independent and measures all changes of address (Courgeau et al. 2013).

Harnessing this finding, Rees et al. (2017) demonstrated algebraically that the slope of the ANMR (measured across multiple levels of scale) is a product of the slope of the CMI and the average MEI measured across different levels of aggregation. Based on this relationship, which was found to hold empirically across a large sample of countries, Rees et al. (2017) proposed the INMI, as a measure which enables robust comparison irrespective of the number of spatial units over which migration is measured. To facilitate cross-national comparison, they advised using the mean across a sample of countries as a benchmark, as follows:

$$INMI = \frac{CMI\ slope\ for\ a\ country}{Average\ CMI\ slope\ for\ all\ countries} * \frac{Mean\ MEI\ for\ a\ country}{Average\ MEI\ for\ all\ countries} \qquad (5)$$

As well as providing a reliable basis for cross-national comparison, the INMI retains the particular advantage of the ANMR in distinguishing the relative contributions of migration intensity and migration effectiveness in generating the aggregate level of population redistribution.

Focusing on countries for which data are available for 20 or more spatial units reduces our sample to 26 countries. We calculated the INMI for these countries and set this index against a global sample of 71 countries including all world regions as reported by Rees et al. (2017). Figure 1 presents a ranking of countries according to their INMI. INMI scores above one indicate the population redistribution impact of internal migration is greater than the global average, with values below 1 denote the opposite. The results indicate that the redistributional impact of internal migration is relatively low in Europe. Of the 27 countries, 20 show internal migration impacts below the global mean, with Spain, Ukraine, Romania and Poland displaying the lowest levels. In contrast, Lithuania and Belarus display the highest levels of population redistribution, more than twice the global mean.

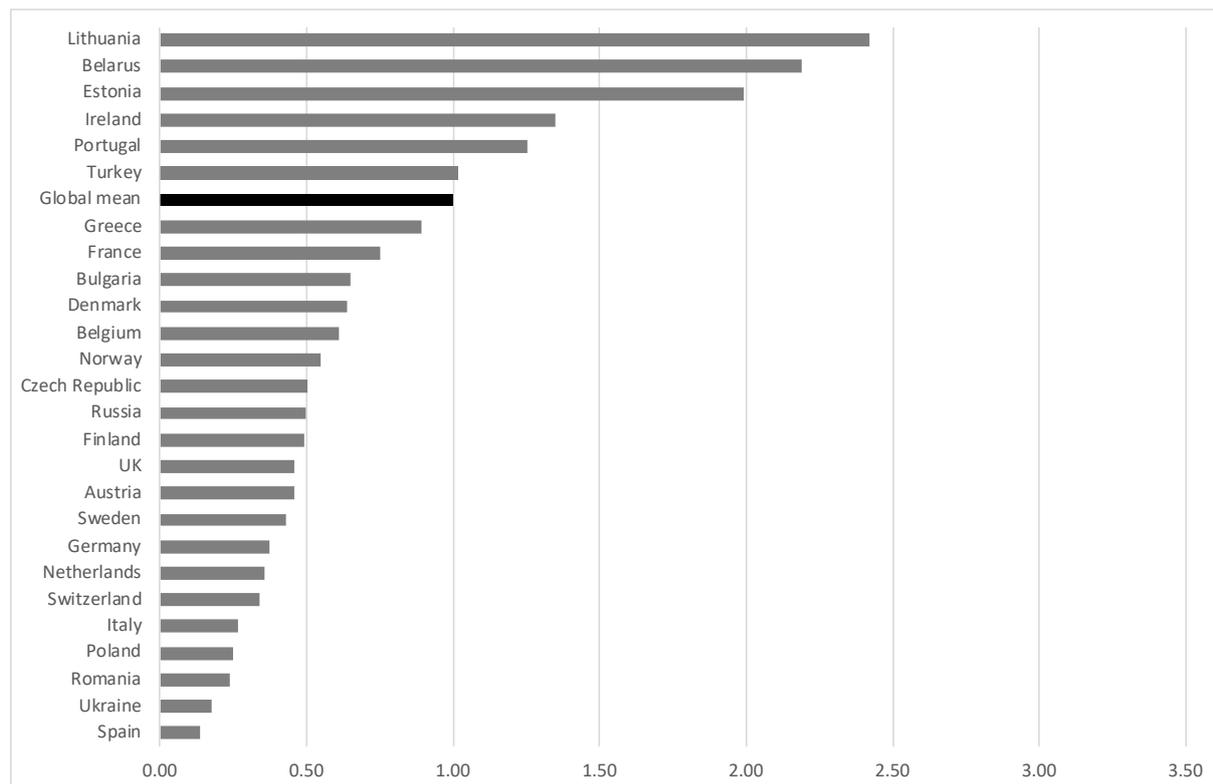

Source: IMAGE Repository, global mean across a sample of 71 countries. Note: Hungary is not included here because an origin-destination matrix is required to estimate the INMI.

**Figure 1** Index of Net Migration Impact

To identify the relative contributions of migration intensity and migration effectiveness to the INMI, Figure 2 reports the standardised ratio of the CMI slope and the average MEI to the global mean. They are denoted C and R, respectively. The plot surface indicates the INMI and the contour curves represent points of equal migration impact according to different combinations of migration intensity and migration effectiveness. The results highlight the complex interaction of these two forces in shaping population redistribution. Belarus stands out with very high levels of redistribution, primarily driven by above average levels of migration effectiveness. Migration effectiveness is 2.5 times the international average and is the principal contributor to the high INMI overall impact experienced in Belarus. For Ireland, Lithuania and Estonia, levels of population of redistribution are also higher than the global mean but underpinned by different mechanisms. In Ireland, both migration intensity and migration effectiveness are above the global average, while in Lithuania and Estonia migration effectiveness is the driving force.

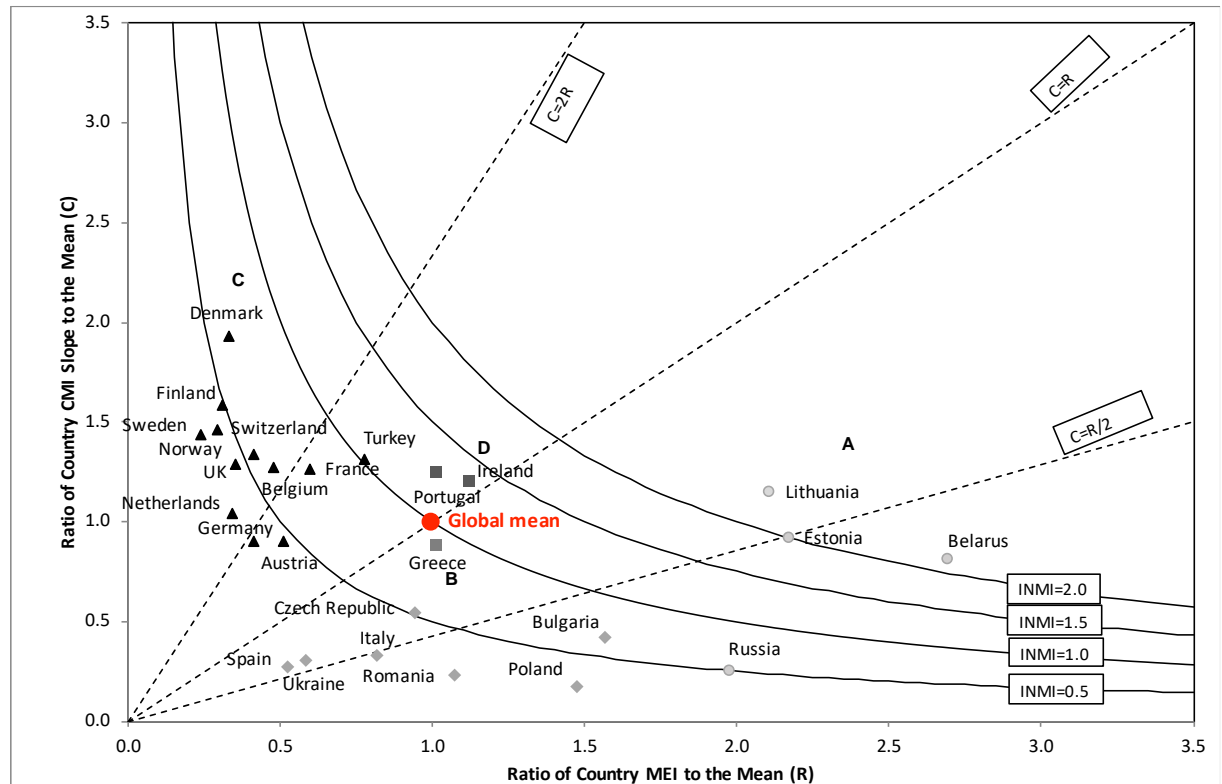

Source: IMAGE Repository, global mean across a sample of 71 countries.
Note: Differing marker colours and shapes indicate the cluster membership based on a kmeans analysis – see text. A=circle, B=Diamond, C=triangle and D=square. Hungary is not included here because an origin-destination matrix is required to estimate the INMI – see Table 1 for details.

**Figure 2** Decomposing the Index of Net Migration Impact: The roles of migration effectiveness and migration intensity.

As Figure 2 shows, four clusters of countries can be identified by performing a k-means cluster analysis on the standardised ratio of the CMI slope and the average MEI to the global mean. Three, four and five clusters solutions were evaluated and a four cluster solution was deemed as the most satisfactory. These clusters are robust to different random-number seeds. Cluster A involves four former member countries of the Soviet Union: Belarus, Estonia, Lithuania and Russia. The main feature differentiating this cluster is very high levels of migration effectiveness with an average standardised MEI ratio of 2.23. In Belarus, Estonia and Lithuania, these very high levels of migration effectiveness are supported by global average levels of migration intensity to produce measures of migration impact which are two times the global average, as indicated by the radial grid in Figure 2. In Russia, however, high levels of migration effectiveness are met by low migration intensity resulting in very modest overall migration impact.

Cluster B encompasses a group of countries in Southern and Eastern Europe, including Italy, Spain and an additional set of former members of the Soviet Union, such as the Czech Republic, Romania, Bulgaria, Ukraine and Poland. In these countries, the average level of migration effectiveness is around the global mean but levels of migration intensity are comparatively low, displaying an average CMI slope ratio of just one third of the global average, resulting in low overall levels of population redistribution. Cluster C involves a large number of Northern and Western countries, including the UK, Scandinavian nations, Belgium, the Netherlands, Germany, Austria, France, Switzerland and Turkey. In this cluster, relatively high levels of migration intensity are absorbed in reciprocal flows, resulting in low migration effectiveness, which constrains the extent of population redistribution. Cluster D includes Greece, Portugal and Ireland where average levels of migration intensity correspond to average levels of migration effectiveness, leading to moderate migration impact.

These patterns constitute a significant finding. As previously demonstrated (Bell et al. 2015b; Rees/Kupiszewski 1999; Rowe 2018a; Sánchez/Andrews 2011), there is a strong spatial gradient of high migration intensity in northern and western countries, moving to low migration levels in southern and eastern European countries. Bell et al. (2015b) found a modest association between the level of economic development and migration intensity across a global sample of countries. This association is present in this study, with the highest intensities recorded in more economically advanced countries of Europe. Economic factors are not, however, sufficient to explain the observed differences. Culture also plays a role. Bernard (2017) revealed a strong association between the timing of departure from the family home and the number of subsequent migrations, affecting overall migration intensities. Early departures from the parental home are common in northern and western Europe, while delayed exits are the norm in countries in southern and eastern Europe.

Our results reveal that the spatial gradient observed for migration intensity dissolves in the case of migration effectiveness. Countries with relatively low national incomes, such as the Baltic States (Cluster A), record the highest MEIs. In contrast, high-income countries in northern and western Europe tend to record lower values (in Cluster C). This pattern is consistent with Rees et al. (2017) which posited that in early stages of development uneven patterns of regional development would trigger high levels of effectiveness, with a return to more symmetrical flows as regional disparities eased. It is important to note that spatial clustering across Europe is not as strong as it was for intensity. This is likely due to the contingencies of national space economies, reflecting, for example, the location of natural endowments, as well as regional policy and international migration patterns. In terms of overall migration impact, high migration effectiveness balances low migration intensity in southern and eastern European countries, while low levels of migration effectiveness offset high migration intensity in northern and western nations.

## 5. Net Internal Migration and Population Density

The system-wide measures used in the previous section indicate the overall impact of migration in redistributing population and help reveal the underlying processes, but provide no insight into their spatial manifestation. Prior work has focused particularly on the contribution of rural-urban migration to the urbanisation process. However, the urban/rural dichotomy is problematic for cross-national comparisons because countries differ in the way they define rural and urban areas. Moreover, the urban/rural dichotomy represents a very coarse classification of space, and few countries classify the rural/urban status of migrants at the start of the migration interval which precludes rigorous analysis. To sidestep these problems, Rees et al. (2017) adopted an alternative approach based on population density. Following Fielding (1989) and Rees/Kupiszewski (1999), this aims to capture a continuum in the settlement hierarchy by classifying administrative areas based on their population density. Building on the ideas originally advanced by Courgeau (1973), area-specific net migration rates are then set against the logarithm of population density for each corresponding area. Application of population-weighted ordinary least squared regression produces an index (the slope of the regression line) which indicates the direction and strength of redistribution across regions. Population-weighted regressions are used to recognise the relative importance of regions within countries and correct for the presence of heteroscedasticity in regression estimates arising from systematic variability in net migration rates because of large variations in population size and outliers (Gujarati 2004). Thus we estimate the following equation:

$$NMR = \alpha + \beta * \log10(den) \quad (6)$$

where NMR represents the region-specific net migration rate; $\log10(den)$ is the logarithm of population density; $\beta$ is the regression slope; and, α is a constant. NMR is calculated as the total migration inflows minus outflows for a region divided by the local total population. Population density is measured by the ratio between the local total population and geographical area in km2. The sign of the slope indicates the direction of the association between NMR and population density, while its value denotes the strength of this relationship. A positive slope denotes net migration gains in more densely populated areas and losses from lower density regions, and points to a process of population concentration (or urbanisation). A negative slope indicates the reverse: net migration losses from high density areas matched by gains in more sparsely populated regions, leading to population deconcentration (or counterurbanisation). In a test across selected countries, the resulting index was shown to be scale independent when calculated for 30 or more spatial units (Rees et al. 2017) so the analysis presented here is confined to the 24 countries in our sample that meet this criterion and for which we have a complete origin-destination matrix

Building on earlier work on the relationship between urbanisation and internal migration by Geyer and colleagues (Geyer 1996; Geyer/Kontuly 1992), Rees et al. (2017) elaborated a conceptual model which proposes a systematic relationship between the patterns of net internal migration and population density as a country progresses through five different phases of development, as displayed in Figure 3. In the first phase, as countries urbanise, a general pattern of net internal migration from low density or rural areas to high density or urban areas is expected. The second phase involves a slight acceleration in the process of urbanisation which is strengthened in the third stage, with internal migration operating to concentrate population in urban areas. In phase 4, this process might reverse into counterurbanisation, or population deconcentration, with net migration flows from more populous areas to less dense regions. The final phase 5 identifies three alternative scenarios: (a) re-urbanisation, (b) counterurbanisation and (c) spatial equilibrium, the latter indicative of no net impact of internal migration on population redistribution. Re-urbanisation may occur where central areas of cities undergo redevelopment. Counterurbanisation may occur as people manifest a preference for low density areas and cities shrink. Alternatively, migration flows across the urban hierarchy may be balanced, resulting in minimal population redistribution, a condition, which Rees et al. (2017) described as a state of spatial equilibrium. Thus, as countries reach higher stages of development, the impact of internal migration on population redistribution is expected to become more limited, and a diversity of outcomes is predicted. At stage 5, the relationship between net migration and population density can fluctuate over time depending on housing and job market conditions.

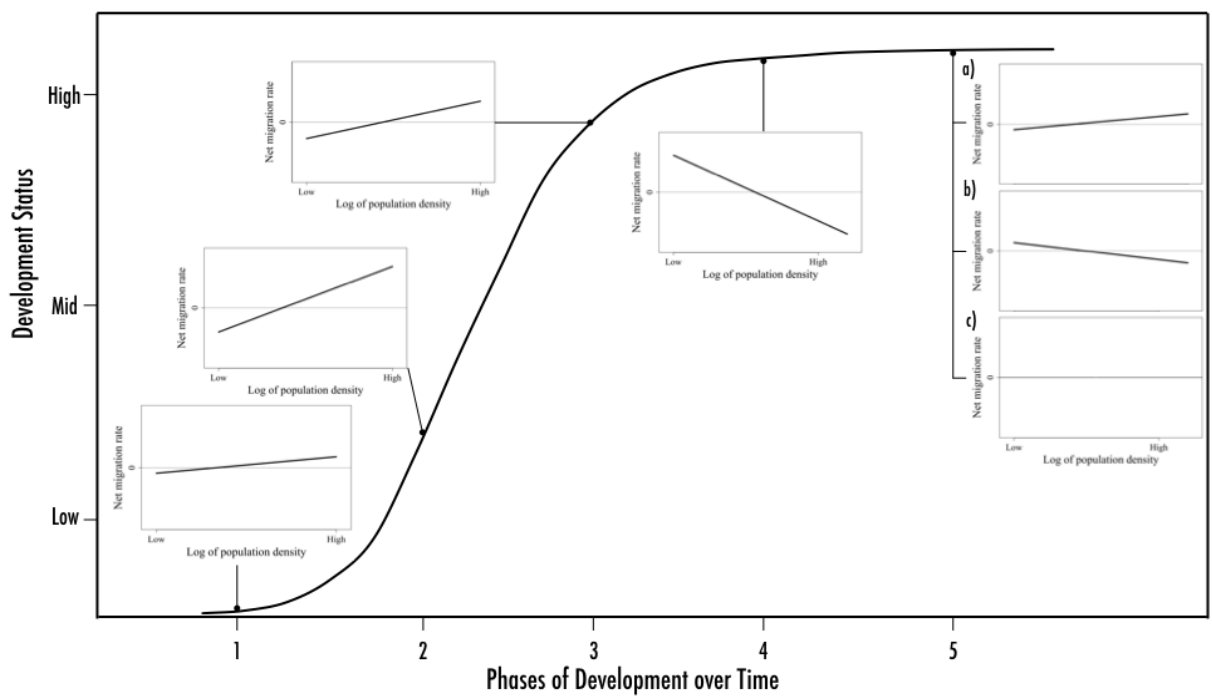

Source: Adapted from Rees et al. (2017), Figure 8.
**Figure 3**. Theoretical framework linking development to population redistribution through net internal migration.

We estimate Equation (6) to analyse patterns of net migration losses and gains. By way of example, Figure 4 plots the relationship between net internal migration rates and population density for Belarus, Lithuania and Belgium, and presents population-weighted estimates of linear regression with 95% confidence intervals. A positive regression slope of 0.73 in Belarus points to a pattern of population concentration resulting from net migration losses in low density regions and gains in densely populated areas, including the districts of Grodno, Pinsk and Viciebsk, which are large regional centres. The Belorussian settlement system stands out by the absence of intermediate cities and the presence of a few urban agglomerations that coexist with a disproportionately large number of low density regions. In our sample, a similar settlement pattern of population settlement is observed only in Romania. A negative slope for Belgium, -0.88 indicates the reverse pattern, net migration gains in low density areas coexisting with losses in more densely settled areas. The steepness of the slope indicates the pace of population deconcentration reflecting processes of suburbanisation and counterurbanisation. At the same time, the large dispersion of regions suggests that while the regression line captures the overall tendency, there are in some countries very large residuals from the regression, which point to the fact that the picture is more complex that density alone predicts. Finally, the flat slope observed in Lithuania as shown by the absence of an association between net migration rates and population density and low adjusted $R_2$ indicates that migration flows across the settlement hierarchy are closely balanced, resulting in minimal population redistribution. Results for all countries can be found in Appendix A.

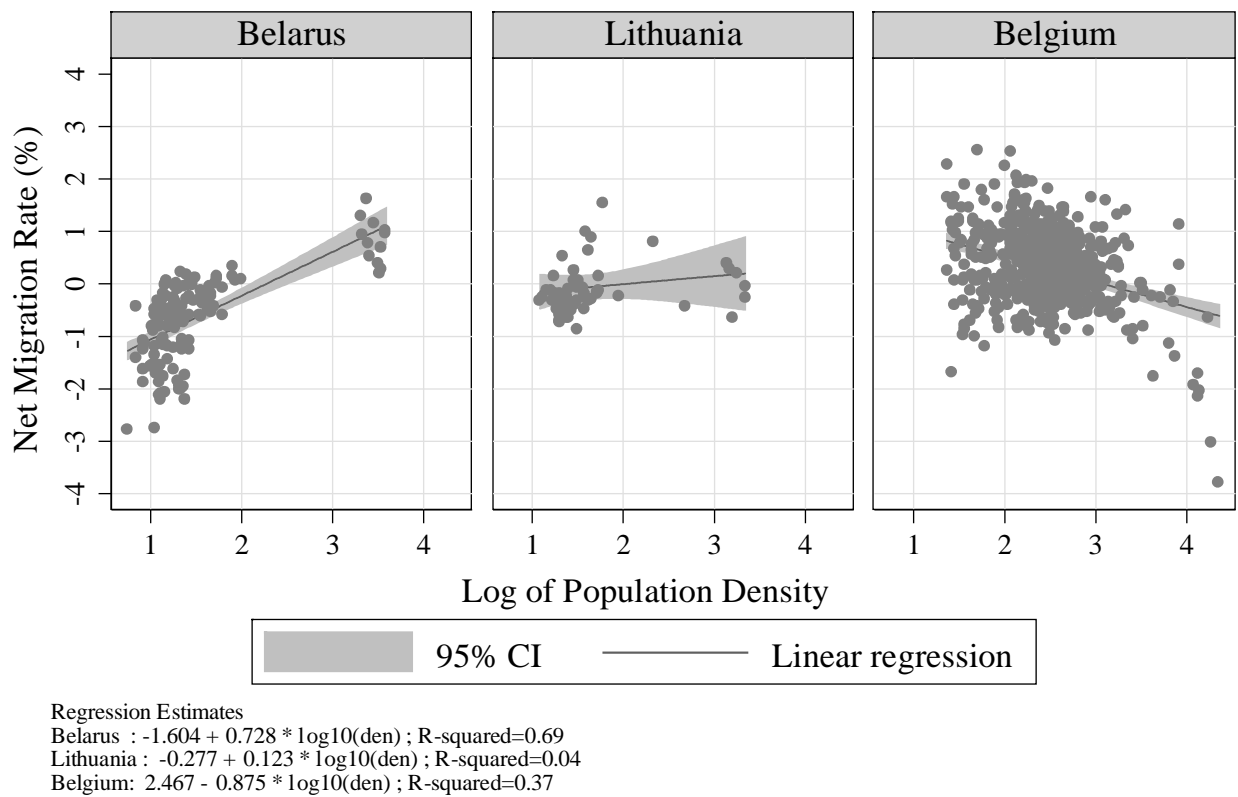

Regression Estimates
Belarus  : -1.604 + 0.728 * log10(den) ; R-squared=0.69
Lithuania :  -0.277 + 0.123 * log10(den) ; R-squared=0.04
Belgium: 2.467 - 0.875 * log10(den) ; R-squared=0.37

Note: Linear regressions were estimated at the administrative level in Table 1, with robust standard errors using the Huber-White sandwich estimator, and were population-weighted to recognise the relative importance of regions within a country. All coefficients reported are significant (p-value<0.05).
**Figure 4**. Relationship between net internal migration rates and the logarithm of population density in Belarus, Lithuania and Belgium.

Table 2 ranks the slopes and reports the corresponding level of significance and adjusted coefficient of determination $R_2$ for 24 countries. It shows that the predominant pattern is a process of population concentration. Although the strength of population redistribution through migration varies, 12 of the 24 countries display slopes which are positive and statistically significant, pointing to net migration gains in densely populated areas and losses in low-density regions. Positive slopes dominate most of Northern, Central and Eastern Europe. To the west and south of the continent, however, this pattern is reversed, with negative slopes across Belgium, Italy and Greece. For these countries the predominant process is one of population deconcentration, whereby gains are occurring in less densely populated areas, fuelled by losses from the more densely populated regions. This pattern is particularly pronounced in Belgium where the slope of -0.88 is driven by population movement to rural municipalities in the Belgium-Luxembourg border, including Martelange, and Léglise, and in the province of Liege, involving Geer and Wasseiges, with corresponding population losses in municipalities within the metropolitan region of Brussels, particularly in Sint-Gillis, Sint-Joost-ten-Node, Schaarbeek and Etterbeek. Six countries - Norway, Spain, Italy, the United Kingdom and two Baltic countries (Lithuania and Estonia) - display slopes which are close to zero and are not statistically significant. These suggest that migration flows across the settlement system are closely balanced, resulting in minimal population redistribution. In the United Kingdom, this pattern represents a major shift reflecting a transition of counterurbanisation fuelled by large net migration losses in London to a pattern of limited redistribution across the country - which is consistent with the pattern of weakening counterurbanisation documented by Lomax/Stillwell (2018). Switzerland also stands out as an unusual case displaying a large positive but statistically insignificant slope and a low $R_2$ value.

These results reflect large migration gains in dense small urban areas - such as in the district of Freienbach and Lac District in the canton of Schwyz and Fribourg respectively - and heavy losses in a small number of remote rural locations in the Goms and Leventina districts but high variability across middle density areas.

**Table 2.** Estimated slopes capturing the relationship between net migration rate and population density

| | Country | Year | Slope | Adjusted $R^2$ |
|---|---|---|---|---|
| Concentration or re-urbanisation | Belarus | 2011 | 0.73*** | 0.69 |
| | Switzerland | 2000 | 0.68 | 0.03 |
| | Portugal | 2011 | 0.65*** | 0.21 |
| | Denmark | 2011 | 0.56*** | 0.63 |
| | Bulgaria | 2006 | 0.49** | 0.27 |
| | Germany | 2009 | 0.41*** | 0.27 |
| | Finland | 2011 | 0.33*** | 0.24 |
| | Netherlands | 2010 | 0.31*** | 0.10 |
| | Russia | 2010 | 0.31*** | 0.58 |
| | Sweden | 2012 | 0.28*** | 0.23 |
| | Turkey | 2012 | 0.27** | 0.03 |
| | Austria | 2010 | 0.22*** | 0.20 |
| | Hungary | 2010 | 0.22** | 0.14 |
| Spatial equilibrium | Norway | 2013 | 0.17 | 0.04 |
| | Lithuania | 2010 | 0.12 | 0.03 |
| | Estonia | 2010 | 0.04 | 0.00 |
| | United Kingdom | 2011 | 0.02 | 0.00 |
| | Spain | 2011 | 0.00 | 0.00 |
| | Italy | 2009 | -0.08 | 0.02 |
| Deconcentration or counterurbanisation | Poland | 2010 | -0.09* | 0.14 |
| | Romania | 2002 | -0.12*** | 0.14 |
| | Czech Republic | 2010 | -0.33*** | 0.09 |
| | Greece | 2011 | -0.52*** | 0.46 |
| | Belgium | 2005 | -0.88*** | 0.37 |

Note: * $p < 0.05$, ** $p < 0.01$, *** $p < 0.001$. Population-weighted regression and robust standard errors were used, based on the Huber-White sandwich estimator (Huber 1967; White 1980). Full estimated regression models are reported in Appendix B.

The model elaborated by Rees et al. (2017) suggests three distinctive trajectories for countries at advanced stages at development: urbanisation, counterurbanisation and spatial equilibrium. Our findings show all three processes are at work within Europe and have operated to form a broad spatial gradient. Population deconcentration and spatial equilibrium patterns spread in the West and South of Europe, while population concentration is a feature shared by many countries in the North, Centre and East of Europe where internal migration gains are observed in urban areas.

At the same time, the results show wide variations within these broad regions. The Czech Republic, Romania and Poland display similar patterns of deconcentration that stand in stark contrast to the predominant pattern of concentration in surrounding countries. Similarly, in Portugal, the Lisbon area has recorded large population gains spurring population concentration which greatly differs from the minimal population redistribution observed in Spain. Behind these variations lies remarkable similarity in the relatively limited impact of internal migration on population redistribution, which is a feature anticipated by Rees et al. (2017) for countries at higher stages of development. Positive slopes in our sample ranges from 0.22 to 0.73, on par with Japan (0.42), but much lower than many developing countries in Asia where China (2.6), Vietnam (3.6), Nepal (4.4) and Mongolia (8.5) are still in the process of the urban transition experiencing substantial population gains in metropolitan areas (Charles-Edwards et al. 2017), and comparable to more developed countries in Latin America, such as Brazil (0.59) and Mexico (0.86) undergoing population concentration (Bernard et al. 2017; Rodríguez-Vignoli/Rowe 2018b).

Figure 5 exposes the spatial structure of net migration gains and losses underpinning this relationship. Here, net migration rates based on administrative areas are superimposed on a base layer showing the major nodes of human settlement, which helps identify the urban areas gaining or losing population from internal migration. Based on country-specific means and standard deviations, standardised net migration rates (z-scores) are reported to help pinpoint areas of unusually high net migration gains or losses; that is, z-scores two standard deviations outside the mean. The results reveal that diverse migration processes underpin the overall processes of population concentration. In Belarus, Portugal, Denmark and Russia, population concentration was driven by unusually high migration gains in a handful of large urban centres, while migration losses were found in sparsely populated areas. Significant gains occurred in the cities of Pinsk, Grodno and Polotsk in Belarus; in the Península de Setúbal within the Lisbon region in Portugal; in the Copenhagen metropolitan area in Denmark; and in Moscow and St Petersburg in Russia. These patterns reflect low variability around the regression line, accompanied by relatively high adjusted $R^2$s as shown in Table 2.

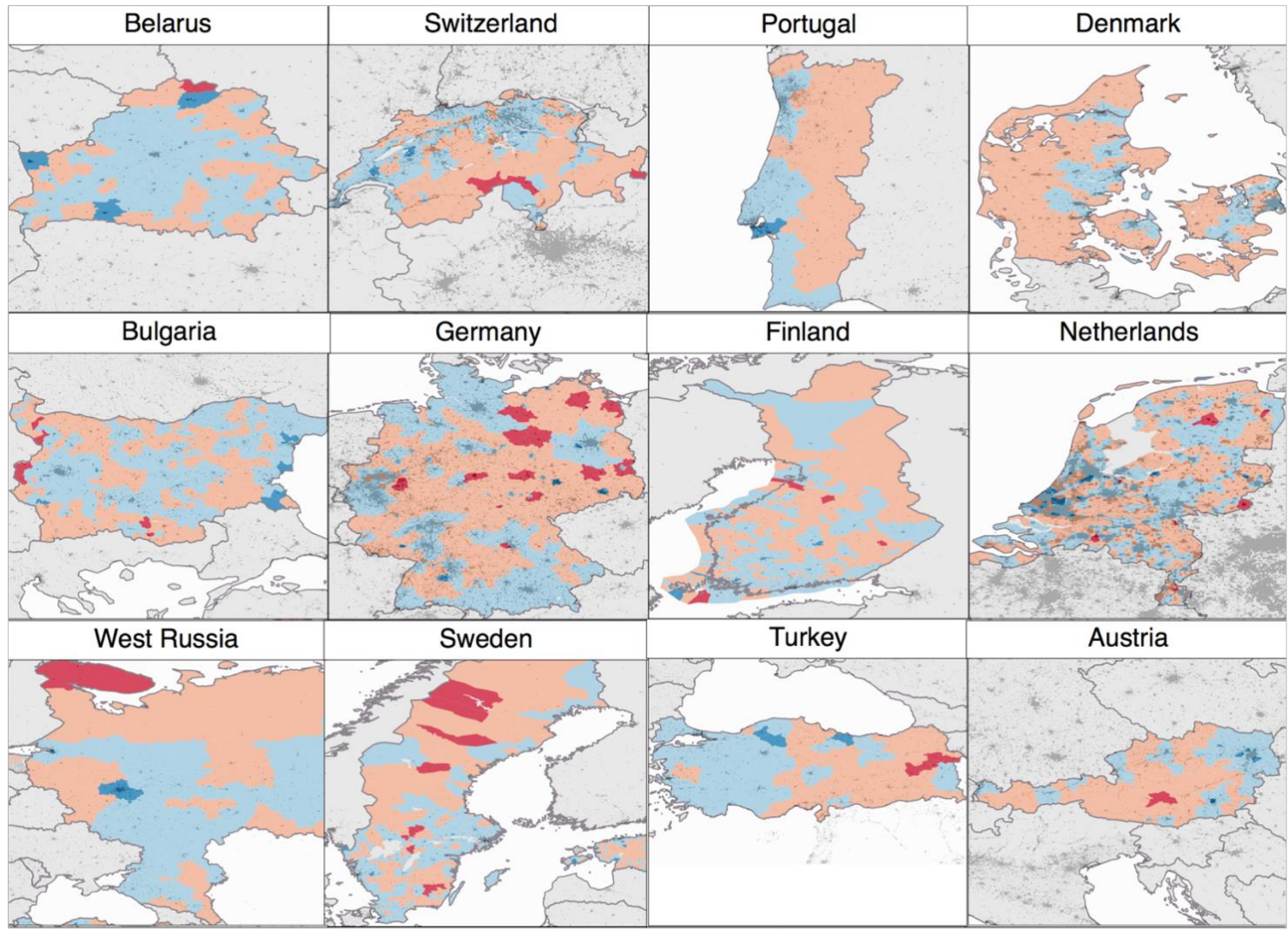

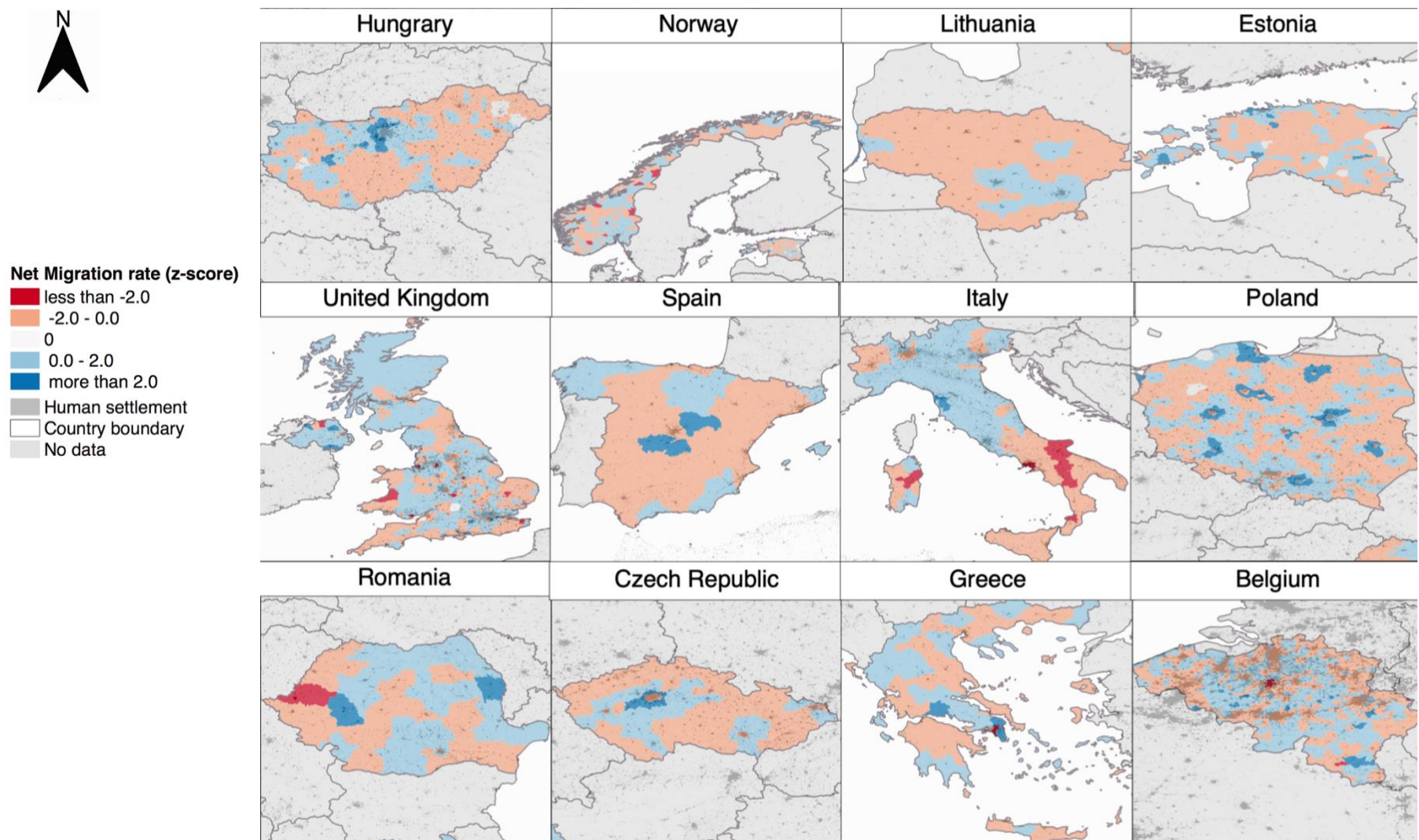

Note: Standardised internal net migration rates are displayed based on country-specific mean and standard deviation. Countries are arranged in the same order as they appeared in Table 2 according to the sign and significance of their net migration slope from statistically significant and positive, through to statistically insignificant and close to zero, to statistically significant and negative.

**Figure 5**. Net internal migration rates (z-scores) for countries with 20 regions or above.

By contrast, patterns of population concentration in Switzerland, Bulgaria, Germany, Finland, the Netherlands, Turkey, Austria, Hungary and Sweden are the result of different processes. In Turkey and Switzerland, significant migration gains in dense small urban areas and heavy losses in a small number of remote rural locations shaped overall patterns of population concentration. In Austria, Bulgaria, Hungary and Sweden, population concentration occurred in large cities, and their satellites, most notably in the areas of Vienna, Sofia, Budapest and Gothenburg. In Germany, on the other hand, migration gains were concentrated in middle-sized and small cities with widespread losses in more remote areas. For example, significant gains occurred in Dresden, Leipzig, Potsdam (near Berlin) and Tübingen (near Stuttgart), while acute losses were observed in Göttingen and around Cottbus in the east of Germany. Patterns of urban sprawl in Germany are largely a response to high property prices in large urban agglomerations (Sander 2018).

Moderate migration gains and losses underpin spatial equilibrium outcomes in Norway, Lithuania, Estonia, Spain, the United Kingdom and Italy, although with higher rates in areas of intermediate population density. In Spain, Estonia, the United Kingdom and Italy, areas of significant gain were located in proximity to key urban centres (eg. Madrid in Spain; Tallinn, Pärnu and Tartu in Estonia; London in the United Kingdom; and, Pisa in Italy), while the cities themselves underwent moderate population losses reflecting local patterns of urban sprawl. Evidence also suggests that this pattern of spatial equilibrium in Italy reflects substitution of internal migration with other forms of mobility, notably international migration and long-distance commuting (Bonifazi et al. 2018). It is notable that Figure 5 reproduces the well-established pattern of migration gains in the north and losses in the south of Italy (Bonifazi/Heins 2000). Norway and Lithuania are distinctive in displaying high migration gains and losses in relatively low density and remote areas.

In the five countries experiencing population deconcentration- Poland, Romania, Czech Republic, Greece and Belgium – net losses have occurred mainly from the larger metropolitan areas. In each case the national capitals - Warsaw, Bucharest, Prague, Athens and Brussels - recorded large migration losses. In Poland, the Czech Republic and Greece, these losses were coupled with strong migration gains on the peripheries reflecting patterns of urban sprawl, rather than a pattern of population reversal (Gordon 1979). In Romania and Belgium, on the other hand, losses from the capital cities were balanced by migration gains in more remote areas scattered across each country.

## 6. Trends in Population Concentration and Deconcentration

As suggested by Rees et al. (2017), the direction and scale of population redistribution through internal migration change in a non-linear fashion as countries develop, reflecting the spatial diffusion of urban development and economic growth. At the upper end of the development spectrum, oscillations may be expected with net migration reinforcing, weakening or reshaping patterns of population concentration or dispersal. As yet however, little is known about the persistence of these patterns. Champion (2001) demonstrated the pervasiveness of counterurbanisation in the British urban system from the 1950s to the 1990s. He also identified marked variations in the intensity of this process, which started in the 1950s and became more pronounced in the 1960s and 1970s during the post-industrial era, with rural areas recording the largest gains in the settlement system. By the 1970s and 1980s the intensity of this process had weakened, reflecting patterns of re-urbanisation coupled with urban sprawl. During the 2000s and 2010s, counterurbanisation continued to weaken with a decrease in magnitude in the net gains and losses recorded in major metropolitan areas and in smaller towns and rural areas, respectively (Lomax/Stillwell 2018) and moved to a pattern of spatial equilibrium (Table 2).

To identify changes in the impact of migration in shaping settlement patterns since the 1980s, Table 3 compares the above findings with those for ten countries reported by Rees/Kupiszewski (1999). Care is needed in comparing results from the two studies because of methodological differences. In particular, Rees/Kupiszewski (1999) did not treat population density as a continuous variable but grouped it into broad categories and examined population density against the rate of population change rather than net migration rate as done in the present paper. Bearing these differences in mind, Table 3 suggests continuing population concentration in Portugal, Germany and Poland and a pattern of stable spatial equilibrium in Italy, but a transition from concentration to spatial equilibrium in Norway and Estonia, and from deconcentration to spatial equilibrium in the UK. In Romania and the Czech Republic, the pattern has reversed from population concentration to dispersal while in the Netherlands dispersal has been replaced by renewed population concentration. It is notable that Rees/Kupiszewski (1999) anticipated the shift to population deconcentration in the Czech Republic. They observed a weak urbanisation process, with areas of medium density gaining migrants from both high and low density areas, reflecting ongoing rural depopulation and local suburbanisation. From the data presented here, this process of urbanisation appears to have halted with population deconcentration now being driven by net internal migration gains in areas of mid-range density.

**Table 3.** Population density and internal migration patterns: 10 countries analysed by Rees/Kupiszewski (1999).

|  | 1980s-1990s* | | | 2000s-2010s** | | |
|---|---|---|---|---|---|---|
| Country | C | SE | D | C | SE | D |
| Portugal | X |  |  | X |  |  |
| Germany |  | X |  | X |  |  |
| Poland | X |  |  | X |  |  |
| Norway | X |  |  |  | X |  |
| Estonia | X |  |  |  | X |  |
| Italy |  | X |  |  | X |  |
| Romania | X |  |  |  |  | X |
| Czech Republic | X |  |  |  |  | X |
| Netherlands |  |  | X | X |  |  |
| United Kingdom |  |  | X |  | X |  |

C: Population concentration; D: Population deconcentration; SE: Spatial equilibrium.

Note: Rees/Kupiszewski (1999) did not explicitly refer to spatial equilibrium but used the term 'intermediate process' to describe countries where most population gains occurred in middle-density areas and population losses in low and high density regions.
Source: * Rees/Kupiszewski (1999), ** results from the present paper

Insights into temporal trends in other countries can also be obtained by comparing the slope of the net migration rates against population density over time. Such analysis is, however, challenging because time-series data are scarce and, even where lengthy time series are available from population registers or censuses, analysis is not straightforward. Comparisons are hindered by changes in administrative boundaries and by the way information is recorded (Rowe 2017), but methods have now been developed to produce temporally consistent spatial frameworks to overcome these problems (Blake et al. 2000; Casado-Díaz et al. 2017; Rowe et al. 2017). Drawing on data from the IMAGE repository, we generated temporally consistent geographies to examine temporal changes in the association between net migration rates and the log of population density in four countries - Finland, Germany, Italy and the Netherlands. These were the only countries for which temporally consistent time series could be assembled.

Figure 6 plots the regression slopes for these countries over recent years, revealing wide variation in both the direction and scale of redistribution across the settlement system from year to year, except in the case of Italy. Italy consistently displays a negative coefficients pointing to the persistence of limited population redistribution with small-scale counterurbanisation or spatial equilibrium with migration losses from high density areas and gains in less populous regions. Finland displays a consistently positive but declining slope from 1995 to 2003, followed by a gradually increasing trend. This suggests a stable but weakening pattern of population concentration, with consistent gains in urban agglomerations and losses from low density regions, transitioning in 2003 to a phase of stronger re-urbanisation. In contrast, Germany displays a steady progression from a moderately negative to a positive slope, reflecting a continuing transition from spatial equilibrium as observed by Rees/Kupiszewski (1999) to concentration arising from internal migration after reunification. At a lower intensity, a similar trend is observed in the Netherlands with a pronounced drop at the start of the 2000s, consistent with the pattern of deconcentration observed by Rees/Kupiszewski (1999) and later transition to a positive slope. This shift to population concentration is thought to reflect the increased clustering of economic activity in the Randstad Region (OECD 2014).

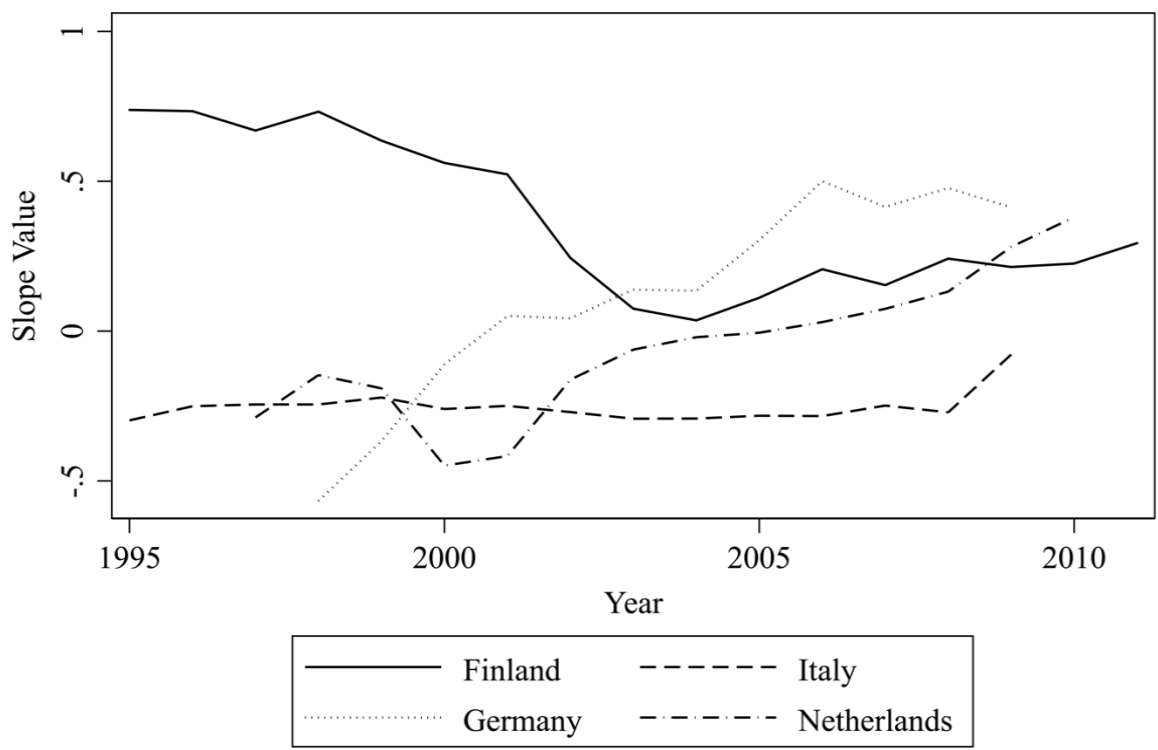

Notes: Linear regressions were estimated using 336 regions for Finland, 412 regions for Germany, 107 regions for Italy and 40 regions for Netherlands. Robust standard errors based on the Huber-White sandwich estimator and population-weighted were used.

**Figure 6**. Fitted slopes capturing the relationship between the net migration rate and population density over time, selected countries.

## 7. Conclusion

Completion of the demographic transition has resulted in migration, both internal and international, becoming the leading agent of demographic change in Europe. While international migration plays an important role in adding population to large urban areas, it makes more limited contributions to populations in regions lower down the urban hierarchy. This underlines the importance of internal migration in transforming settlement systems, particularly in terms of population concentration and deconcentration. However, remarkably little progress has been made in understanding the spatial impact of internal migration in Europe, mainly because of issues of data comparability, coarse dichotomies between rural and urban areas, and the absence of robust comparative metrics. In this paper, we sought to address this gap by harnessing a unique international dataset of country-specific internal migration flow matrices, developed as part of the IMAGE project, and employing a newly developed suite of scale-independent migration indicators. Using the Index of Net Migration Impact (INMI), we first quantified the impact of migration on population redistribution at a national level in 26 countries. We decomposed the INMI into its constituent elements to determine the influence of migration intensity and migration effectiveness on the resulting level of migration impact. In 24 of the 27 countries in our sample, we then examined regional net migration rates to assess the spatial effects of this redistribution, regressing net migration against the log of population density, as a proxy for urbanisation.

Our results reveal that the effect of internal migration on population redistribution, as measured by the INMI, is relatively low across Europe, with more than 80 per cent of countries in our sample showing a redistributive effect below the global average. This was particularly marked in Spain, Ukraine, Romania and Poland. With the exception of Lithuania and Belarus, which show levels of population redistribution more than twice the global mean, there is limited variation between European countries in the redistributive effect of internal migration. However, decomposition of the INMI into its constituent elements revealed that the intensity and effectiveness of migration follows a marked spatial gradient. In northern and western Europe, the low redistributive effect of migration is the result of high migration intensity being absorbed by reciprocal migration flows, whereas in southern and eastern Europe migration flows are highly imbalanced but their effect on population redistribution is offset by low migration intensity. The presence of a clear spatial gradient of high mobility in the north and west of Europe moderating toward the South and East is consistent with prior studies, but our results demonstrated that this spatial gradient dissolves in the case of migration impact because migration effectiveness and migration intensities vary in an inverse manner across Europe. Despite broad similarities in the overall impact of migration in redistributing population, there are significant variations between countries in its effects on the settlement pattern. In the north and east of the continent, population redistribution is directed toward more densely populated regions, leading to population concentration or re-urbanisation, whereas this pattern is reversed in the south and west where population gains are focused on lower-density regions, contributing to population deconcentration or counterurbanisation. This broad spatial gradient is interrupted by a number of countries which have reached a spatial equilibrium under which internal migration alters the existing patterns of population settlement only minimally, as is the case in Norway, Lithuania, Estonia, Spain, the United Kingdom and Italy.

These findings lend support to the model advanced by Rees et al. (2017), which hypothesised diverse trajectories for countries at the upper end of development ladder, contrasting with earlier phases where migration gains are systematically directed toward more densely populated regions. Our time-series analysis confirmed variability between countries but also revealed variation over time, with an overall trend broadly toward increased population concentration in the four European countries in our sample. In Germany and the Netherlands, the early 2000s saw a shift from population spatial equilibrium to concentration, whereas Finland displayed a persistent pattern of population concentration and Italy continued to experience small-scale counterurbanisation (or spatial equilibrium). These findings demonstrate the important variations that existing within and between the regions of Europe in population redistribution processes while highlighting the diminishing impact of migration in shaping settlement patterns in highly urbanised countries.

These findings underline the importance of simple yet robust indices to tease out commonalities and differences in migration processes between countries. This global overview provides the foundation for further in-depth analysis of the particular patterns of population redistribution within countries as Ravenstein pioneered 130 years ago. Since then, the direction of migration flows had evolved, with urbanisation being increasingly replaced by deconcentration and spatial equilibrium in many European countries. These diverging processes confirm the importance of considering flows and counter-flows as originally suggested by Ravenstein in his fourth law (Ravenstein 1885: 199), while highlighting the relevance of conceptualising migration as a system-wide process which extends across the entire settlement system. As shown in this paper, progress in data collection practice and in methods of analysis have permitted a rigorous search for empirical regularities, among which the limited redistributive effect of migration appears to be a fundamental underlying similarity shared by most European countries. Perhaps more important than commonalities are singularities that differentiate countries: the wide variations found in the effect of migration on settlement patterns confirm the need for nuanced investigations at a country-specific level. The country maps and scatter plots of net migration rates included in this paper offer a starting point to select outlying regions that represent particular cases where population gains (or losses) are higher (or lower) that the regression analysis suggests. New methods (Rodríguez-Vignoli/Rowe 2018a, 2018b, 2017) and analytical software (Rowe et al. 2019) have been developed to extend the analysis beyond exploring net migration balances and measure their impacts on the compositional population structure of areas. Additionally, the comparative analysis of migration processes at a local scale would help unravel the social, economic demographic factors that underpin the heterogeneous population dynamics across the countries of Europe.

**Acknowledgements**
The views expressed are those of the authors and are not necessarily those of the United Nations. The work reported in this article forms part of the "Understanding the declining trend in internal migration in Europe" funded by Regional Studies Association under the Early Career Research research grant, and it was also supported under Discovery Project DP110101363, Comparing internal migration in countries around the world: measures, theories and policy dimensions, funded by the Australian Research Council. The article draws from several sources, including the IPUMS database maintained by the Minnesota Population Center and national statistical offices. We gratefully acknowledge the helpful comments from two anonymous referees and the editor Prof. Phil Rees on an earlier version of the paper as well as attendees at the European Regional Science Association conference in Groningen, 2017 and a seminar at the School of Social Statistics within the University of Manchester.

**Appendix A.** Slope of net internal migration rate (z-scores) as a function of log population density

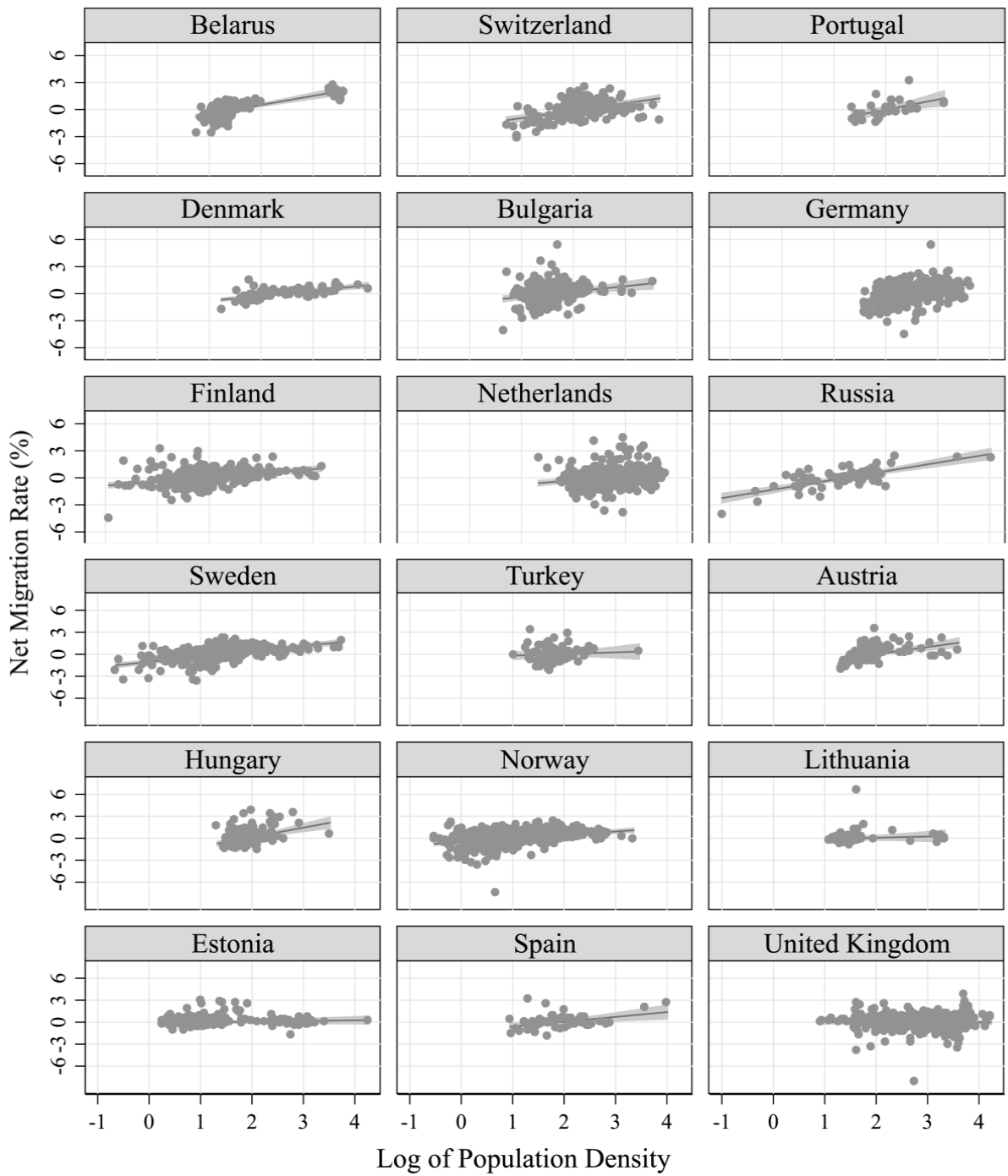

*Submitted to Comparative Population Studies*

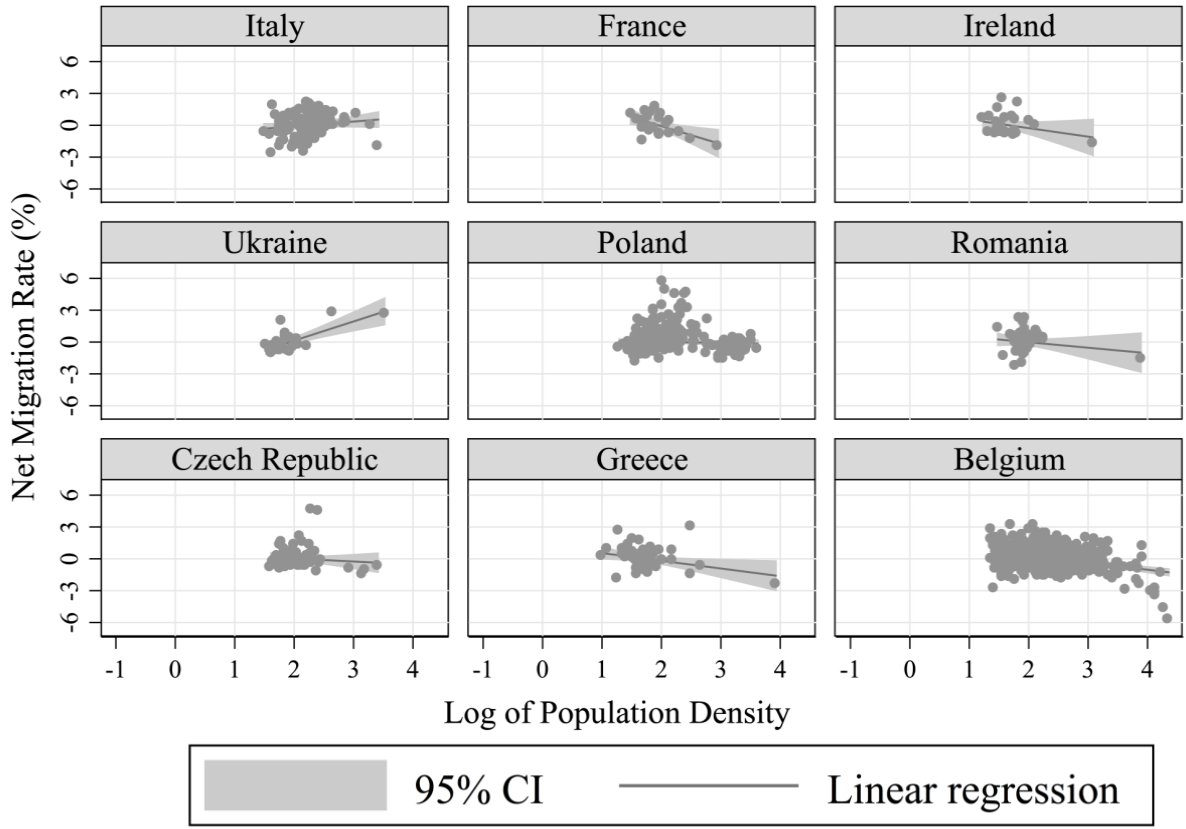

Note: Linear regressions were estimated at the administrative level in Table 1, with robust standard errors using the Huber-White sandwich estimator (Huber 1967; White 1980), and population-weighted to recognise the relative importance of regions within a country. See Appendix B for the estimated regression coefficients.

**Appendix B.** Regression Coefficients between net migration rate and population density.

| Country | Log10(den) | | Constant | | Observations | Adjusted $R^2$ |
|---|---|---|---|---|---|---|
| | Coefficient | t | Coefficient | t | | |
| Belarus | 0.728*** | -12.62 | -1.604*** | (-14.34) | 130 | 0.689 |
| Switzerland | 0.678 | -1.38 | -1.736 | (-1.48) | 184 | 0.03 |
| Portugal | 0.654*** | -4.44 | -1.534*** | (-4.01) | 30 | 0.206 |
| Denmark | 0.555*** | -12.8 | -1.367*** | (-11.99) | 99 | 0.633 |
| Bulgaria | 0.491*** | -3.62 | -1.072*** | (-4.49) | 264 | 0.274 |
| Germany | 0.413*** | -10.65 | -1.085*** | (-10.89) | 412 | 0.268 |
| Finland | 0.333*** | -9.12 | -0.597*** | (-8.48) | 336 | 0.242 |
| Netherlands | 0.310*** | -6.00 | -0.912*** | (-6.48) | 431 | 0.095 |
| Russia | 0.307*** | -9.17 | -0.503*** | (-8.86) | 80 | 0.582 |
| Sweden | 0.275*** | -6.88 | -0.574*** | (-7.24) | 290 | 0.232 |
| Turkey | 0.273** | -2.67 | -0.621* | (-2.17) | 81 | 0.031 |
| Austria | 0.215** | -3.31 | -0.520*** | (-3.76) | 99 | 0.2 |
| Hungary | 0.220** | -2.85 | -0.507** | (-3.25) | 196 | 0.135 |
| Norway | 0.172 | -1.87 | -0.350* | (-2.44) | 428 | 0.039 |
| Lithuania | 0.124 | -1.26 | -0.277 | (-1.34) | 60 | 0.028 |
| Estonia | 0.04 | -0.37 | -0.103 | (-0.36) | 225 | 0.000 |
| Spain | 0.002 | -0.06 | -0.005 | (-0.05) | 52 | 0.000 |
| United Kingdom | 0.02 | -2.67 | -0.054 | (-0.37) | 404 | 0.000 |
| Italy | -0.077 | (-0.75) | 0.19 | -0.79 | 107 | 0.017 |
| Poland | -0.090* | (-2.00) | 0.217* | -2.05 | 379 | 0.014 |
| Romania | -0.115*** | (-4.75) | 0.243*** | -3.65 | 42 | 0.141 |
| Czech Republic | -0.332*** | (-5.26) | 0.779*** | -4.95 | 77 | 0.092 |
| Greece | -0.520*** | (-10.55) | 1.257*** | -8.77 | 54 | 0.463 |
| Belgium | -0.875*** | (-8.24) | 2.467*** | -8.97 | 589 | 0.372 |

Note: *t* statistics in parentheses. * $p < 0.05$, ** $p < 0.01$, *** $p < 0.001$. Population-weighted regression and robust standard errors were used, based on the Huber-White sandwich estimator